\documentclass{article}
\usepackage{graphicx} 
\usepackage{amsmath} 
\usepackage{amssymb}
\usepackage{biblatex}
\usepackage{tabularx}
\usepackage{algorithm2e}
\usepackage{makecell}
\usepackage{arxiv}

\title{SignedLouvain: Louvain for signed networks}

\author{
John N. Pougué-Biyong \\
Mathematical Institute \\
University of Oxford \\
\texttt{john.pougue-biyong@maths.ox.ac.uk} \\
\And
Renaud Lambiotte \\
Mathematical Institute \\
University of Oxford \\
\texttt{renaud.lambiotte@maths.ox.ac.uk} \\
}

\date{July 2024}

\begin{document}
\maketitle

\begin{abstract}
 In this article, we consider the problem of community detection in signed networks.
 We propose \texttt{SignedLouvain}, an adaptation of the Louvain method to maximise signed modularity, efficiently taking advantage of the structure induced by signed relations. We begin by identifying the inherent limitations of applying the standard Louvain algorithm to signed networks,   before introducing a novel variant specifically engineered to overcome these challenges. Through extensive experiments on real-world datasets, we demonstrate that the proposed method not only maintains the speed and scalability of its predecessor but also significantly enhances accuracy in detecting communities within signed networks. 
\end{abstract}

\section{Introduction}

Community detection in networks has emerged as a fundamental task in the study of complex systems \cite{fortunato2016community,schaub2017many}. It involves the identification of densely connected groups of nodes, often called communities or clusters, within a larger network. These communities often represent functional units, such as groups of friends in social networks, related proteins in biological networks, or coherently connected web pages in the World Wide Web \cite{girvan2002community,ahn2010link,leskovec2010empirical}. Uncovering communities is crucial for understanding the structure and function of networks, for instance to infer hidden patterns in social and biological systems but also to identify structural patterns hindering diffusion on networks \cite{lambiotte2021modularity}. The ability to accurately identify  communities enables researchers and practitioners to gain deeper insights into the intrinsic properties of networks, in various domains ranging from sociology to computer science and economics.

Signed networks introduce an additional layer of complexity into the study of networked systems by incorporating both positive and negative relationships between nodes \cite{kunegis2010spectral}. Unlike traditional networks where links typically signify a positive association or similarity, signed networks explicitly encode both friendly (positive) and antagonistic (negative) interactions. This is representative of social systems where relationships can be cooperative or adversarial, as well as ecological systems where species may exhibit symbiotic or competitive interactions \cite{szell2010multirelational,leskovec2010signed,symeonidis2013biological}. The presence of negative edges challenges conventional community detection methods, as they must now discern not only cohesive groups of nodes but also account for the divisive relations at play \cite{traag2009}. Moreover, the resulting communities are expected to arise from other mechanisms than homophily \cite{mcpherson2001birds}, for instance via different forms of structural balance \cite{harary1953notion,davis1967clustering}. In general, understanding and accurately partitioning signed networks is pivotal for a comprehensive analysis of such systems, enabling us to capture the nuanced interplay of cooperative and antagonistic relationships within a network.

In this article, we consider the problem of community detection in signed networks, and investigate heuristics to maximise the signed modularity function \cite{traag2009}. 
Our purpose is to study generalisations of the Louvain method~\cite{blondel2008fast}, a popular and efficient method for modularity optimisation in unsigned networks. We first show the bottlenecks of trying to apply the standard Louvain algorithm on signed networks and propose {SignedLouvain}, an efficient adaptation of the Louvain method specifically designed for signed networks. Through real-world and synthetic experiments, we show that {SignedLouvain} combines is an ideal trade-off between speed and accuracy. 
\section{Louvain on signed networks}

In Section~\ref{subsec:louvain_approach}, we describe the Louvain algorithm~\cite{blondel2008fast,blondel2023fast} and how it has been extended to be applied on signed networks. In Section~\ref{subsec:louvain_performance}, we evaluate the performance and caveats of the approach on signed networks.

\subsection{Approach} \label{subsec:louvain_approach}
Given a signed, undirected, weighted network $G$, with $n$ nodes, we note $A \in \mathbb{R}^{n \times n}$ its adjacency matrix. We decompose $ A = A^{+} - A^{-} $ where $A^{+}$ and $A^{-}$ have non-negative entries and represent the positive and negative parts of the network respectively.

Given this network specification, the Louvain algorithm~\cite{blondel2008fast} detects communities on unsigned networks ($A^{-} = 0$) using a modularity maximisation-based approach. The modularity of an unsigned network partition is a scalar value between $-1$ and $1$ that measures the density of links (edges) inside communities as compared to its expectation with respect to a null model. Based on the homophily assumption, the intuition is that a well partitioned network should have most links inside communities and not between communities. Given an unsigned network with adjacency matrix $A$, the (generalised) modularity of partition $\sigma$ is defined as:
\begin{align} \label{eq:modularity_def}
    Q(\{\sigma\}) &= \frac{1}{2m} \sum_{i,j} \left(A_{ij}- \gamma \frac{k_i k_j}{2m}\right) \delta(c_i, c_j)
\end{align}
where $A_{ij}$ is the edge weight between $i$ and $j$, $k_i = \sum_j A_{ij}$ is the weighted degree of node $i$, $c_i$ is the community to which node $i$ is assigned, $m$ is the sum of all edge weights, $\gamma$ is a resolution parameter \cite{reichardt2006statistical,lambiotte2014random} - the original expression for modularity is recovered when $\gamma=1$ \cite{girvan2002community} - , and $ \delta$-function is the indicator function.

Finding the partition that maximises Equation~(\ref{eq:modularity_def}) is a NP-hard problem \cite{brandes2007modularity}, so different heuristics have been proposed for its optimisation.  A popular heuristic algorithm is the so-called Louvain method \cite{blondel2008fast}. Starting with a partition where each node is in its own community, it alternates between two phases until convergence: (1) a \textit{move} phase where each node is moved in the neighbouring community (i.e. community of its neighbours) that maximises modularity increase, (2) once no move increases modularity anymore, an \textit{aggregate} phase consisting in building a new network whose nodes are now the communities found during the first phase, and edges are the sum of link weights between these communities. It iterates between the two steps until no improvement on modularity is achieved. 
The method is known to show a good balance between accuracy and speed of execution. The speed of the method is ensured by the locality of each operation, as vertices only move to neighboring communities and  calculating the resulting change of modularity  is a cheap operation. Its accuracy arises from its flexibility, as vertices may be removed from their community and re-assigned to others in later stages, and its multi-scale nature, as the move phase is first done locally, and then over longer distance as the vertices are aggregated. 
One bottleneck is that the method may converge to a local maxima which is not a global maxima. It is therefore preferable to run the method several times to find the most optimal configuration \cite{lambiotte2014random}.

Originally proposed for unweighted, undirected networks, the modularity function has been extended to more general classes of networks, together with the algorithms required for their optimisation, including Louvain \cite{blondel2023fast}. In the case of signed networks, a signed modularity function has been defined~\cite{traag2009} as: 
\begin{align} \label{eq:signed_modularity_def}
     Q(\{\sigma\}) &= \frac{1}{2m} \sum_{i,j} \left[A_{ij} - \left( \gamma^{+} \frac{k^{+}_i k^{+}_j}{2m^{+}} - \gamma^{-} \frac{k^{-}_i k^{-}_j}{2m^{-}} \right) \right] \delta(c_i, c_j)
\end{align}
where $k^{\pm}_i = \sum_j A^{\pm}_{ij}$ is the weighted degree of node $i$ on the positive (resp. negative) part of the network,  $\gamma^{\pm}$ are resolution parameters, and $m^{\pm}$ is the sum of all positive (resp. negative) edge weights. 
This definition of modularity aims to reward configurations with positive internal links and absent internal negative links, and penalise configurations with negative internal links and absent internal positive links. 
In practice, this function can also be maximised with the original Louvain algorithm\footnote{\url{https://louvain-igraph.readthedocs.io/en/latest/}} or related extensions, as we discuss more in detail below.

\subsection{Caveats} \label{subsec:louvain_performance}

The presence of negative connections in the network increases the existence of local maxima when optimising modularity Q (Eq.~\ref{eq:signed_modularity_def}). Consider $\Delta Q$ the change of modularity occurring from moving an isolated node $i$ into a community $C$. There are network structures and edge densities for which $\Delta Q$ is negative for any possible move. A simple example shows that this can even occur at the start of the optimisation process: take a signed network with 5 nodes and 4 negative edges connecting node 0 to nodes 1, 2, 3 and 4. Following the Louvain algorithm, the initial partition is made of singletons. The variation in modularity occurring when moving any node in a neighbouring community is:
\begin{align}
\Delta Q &= \frac{1}{m} \left( A_{ij} + \gamma^{-} \frac{k^{-}_i k^{-}_j}{2m^{-}} \right) = \frac{1}{4} \left( -1 + \frac{1}{2}\gamma^{-} \right) < 0
\end{align}
for common values of $\gamma$, i.e. $ \forall \text{ } \gamma^{-} \in \left[0, 2 \right]$. As a consequence, the final configuration of the optimisation process is the initial configuration: $\sigma_0 = \{ \{0\}, \{1\}, \{2\}, \{3\}, \{4\} \}$. What final configuration should we expect? In a balanced signed network \cite{harary1953notion}, in virtue of the strong structural balance assumption (\textit{the enemy of my enemy is my friend}), we expect the optimal configuration to be $\sigma = \{ \{0\}, \{1, 2, 3, 4\} \}$. This can only occur if nodes are able to move into the communities of non-neighbouring nodes, which is not permitted in the original Louvain algorithm. An extension of the Louvain algorithm, that we denote RelaxedLouvain from now on\footnote{\url{https://louvain-igraph.readthedocs.io/en/latest/}}, allows nodes to move into any community, whether any neighbouring node belongs to this community or not. In essence, this means that the optimisation process explores significantly more configurations to find the modularity-maximising one. In that setting,  the optimisation algorithm terminates at partition $\sigma_f = \{ \{0\}, \{1, 2, 3, 4\} \}$. Numerically, for $\gamma^{-} = 1$, we confirm that maximum modularity is achieved with $\sigma_f$ ($Q = 0.5$) and underachieved with $\sigma_0$ ($Q = 0.3125$).

In the following experiment, we show that the neighbourhood restriction imposed in the Louvain algorithm makes it converge towards partitions with plenty of small communities (often singletons) in sparse networks, with and without a preponderant number of negative edges. This type of configurations is not realistic in most real-world cases, e.g., in opinion networks or online debates where only a few major opinions exist. It is also contrary to the purpose of community detection, i.e. finding meso-scale representations of networks. RelaxedLouvain is a way to lift this restriction and approximate the global maxima more closely. We consider a standard Signed Stochastic Signed Model (SSBM)~\cite{holland1983stochastic,jiang2015sbm} with 3 communities of size 30, 20, and 10, where all intra- (inter-) community edges are positive (negative). We denote $p_{in}$ and $p_{out}$ the share of existing positive and negative edges respectively. We vary $p_{in}$ and $p_{out}$ between $80\%$ and $0\%$ and generate random networks repeatedly. For each generated network, we run Louvain and RelaxedLouvain, and compute the Normalised Mutual Information (NMI)~\cite{strehl2002cluster, fortunato2016community} of their final partition with the planted one. The NMI is averaged across all the runs of a given couple $(p_{in}, p_{out})$.

\begin{figure}[ht]
\includegraphics[width = \textwidth]{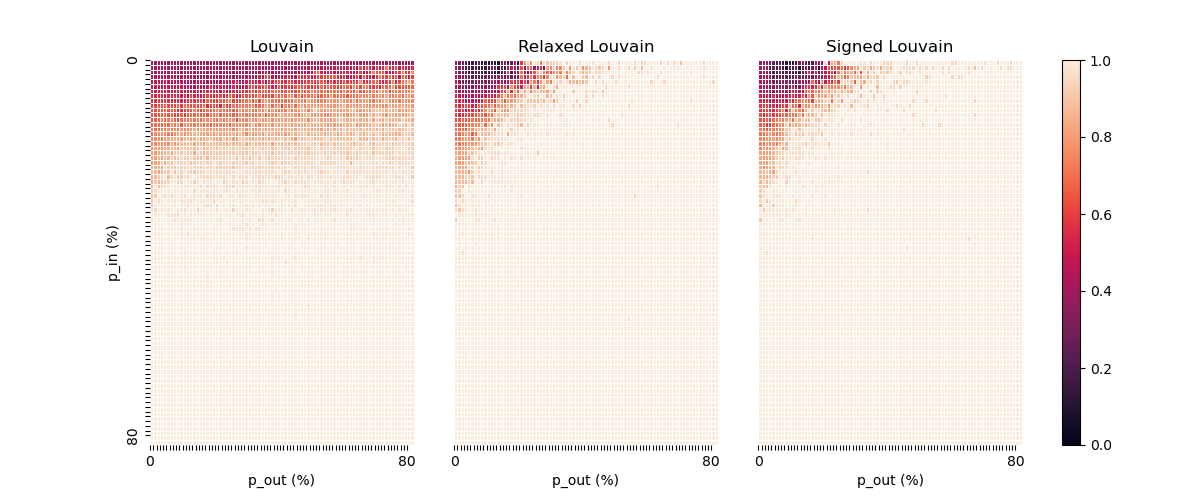}
\caption{Performance of Louvain v RelaxedLouvain v Signed Louvain on a Signed Stochastic Block Model. The colour indicates the NMI of the final partition with the one used to build the SSBM. As can be seen, the quality of the original Louvain remains low in situations when the density of negative edges between communities is large.}
\label{fig:LvRLvSL}
\end{figure}

Figure~\ref{fig:LvRLvSL} depicts the results of the experiment. We observe the over-reliance of the Louvain algorithm on positive edges such that when the proportion of such edges is low ($p_{in} < 20\%$), the performance of the method dramatically decreases. A closer inspection shows that Louvain converges towards partitions composed of many small groups (2-5 nodes) when $p_{in}$ is low. On the other hand, RelaxedLouvain is able to recover the true community structure even when $p_{in}$ is low and $p_{out}$ is higher, at the only condition that the network is not too sparse ($p_{in}$ and $p_{out}$ close to 0).

\begin{figure}[ht]
\includegraphics[width = \textwidth]{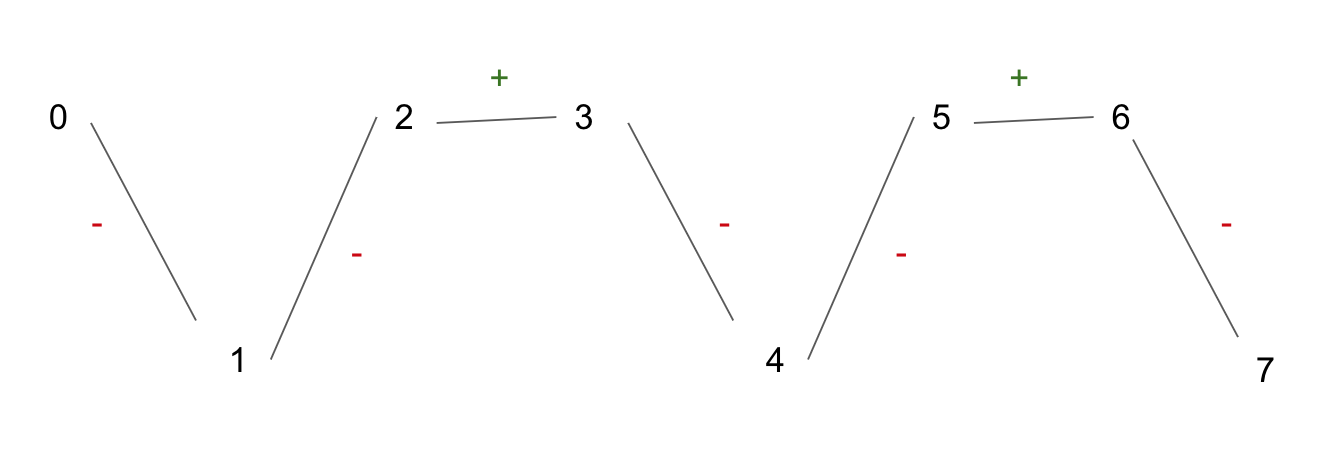}
\caption{Example of graph where  RelaxedLouvain may produce inconsistent clusters, for instance $ \{0, 5, 6\}$, due to to its lack of locality during the optimisation.}
\label{fig:proximity}
\end{figure}

Even though RelaxedLouvain seems to be a promising alternative to Louvain to unveil communities in signed networks, it also suffers from limitations. The first caveat with RelaxedLouvain is the computational cost: scanning every community for every node at every optimisation step increases dramatically the number of operations involved. It is prohibitive in real-world settings where networks are much larger than the synthetic ones considered in the previous example. Moreover, scanning all communities without taking into consideration their proximity to the current node may lead to inconsistent community structures. Indeed, in the example shown in Figure~\ref{fig:proximity}, from the perspective of structural balance, node 0 can be grouped with nodes 2 and 3, node 3 can be grouped with nodes 5 and 6, but to ensure  consistency on the distance between nodes in the same community, node 0 should not be grouped with nodes 5 and 6, without being grouped with nodes 2 and 3 first. However,  RelaxedLouvain converges towards configurations such as $\sigma = \{ \{0, 5, 6\}, \{2, 3 \}, \{1, 4, 7 \} \}$ because it ignores the proximity in the network and randomly select communities to which a given node can be assigned. In larger networks, this may bring severe consistency issues where nodes can potentially be in the same community as nodes that cannot even be reached via a  path.

This behaviour is not expected in unsigned graphs because moving singleton into the community of a non-neighbouring node cannot increase modularity. However, in the presence of a negative null model, a singleton can join another singleton even if they are not connected. To illustrate this, let us take two non-connected nodes $i$ and $j$ and look at their pairwise term in the modularity:
\begin{align}
M_{ij} &= \frac{1}{m}\left[A_{ij} - \left( \gamma^{+} \frac{k^{+}_i k^{+}_j}{2m^{+}} - \gamma^{-} \frac{k^{-}_i k^{-}_j}{2m^{-}} \right) \right] \delta(c_i, c_j) \\
    &=  \frac{1}{m}\left( \gamma^{-} \frac{k^{-}_i k^{-}_j}{2m^{-}} - \gamma^{+} \frac{k^{+}_i k^{+}_j}{2m^{+}} \right) \delta(c_i, c_j) \label{eq:mij}
\end{align}
as $A_{ij} = 0$. In that case, the modularity gain depends on the relative contribution of the two concurrent null models (positive and negative). In particular, the gain is positive when the negative null model has more contribution for these two nodes. In the example shown in Figure~\ref{fig:proximity}, node $0$ has no positive edge so $k^{+}_0 = 0$ and only the first term in Equation~(\ref{eq:mij}) is non-zero, hence $M_{ij} > 0$.

\section{SignedLouvain: Louvain for signed networks} \label{sec:signedlouvain}

In this section, we present {SignedLouvain}, our proposed extension of the Louvain method \textbf{for} signed networks. Unlike RelaxedLouvain, it relaxes the neighbourhood restriction imposed by Louvain while taking into account the proximity in the network. We will show, through real-world experiments, that {SignedLouvain} offers a good tradeoff between speed, consistency and accuracy.

\subsection{Method} \label{subsec:signedlouvain_approach}
{SignedLouvain} aims to optimise Eq.~\ref{eq:signed_modularity_def} by adopting a multiplex network approach \cite{kivela2014multilayer} where different moves are allowed or forbidden depending on the layer the current node operates in. In this case, the layers are the positive and negative parts of the network. We define $d_{+}$ and $d_{-}$ the number of hops  a node is allowed to move on the positive and negative network to adopt a new community. At the \textit{move} step, the algorithm skims all the communities within $d_{+}$ hops and $d_{-}$ hops from the current node and selects the one maximising modularity locally. 

In order to find the best values for $d_{+}$  and $d_{-}$, it is important to carefully identify which nodes are likely to be in the same community than the target node that one tries to move.
 Ideally, the set of these nodes should be as small as possible, to accelerate the algorithm, but not too small, to avoid losing precision. In the case of unsigned network, the archetype of a community structure would be a system composed of multiple connected components. The latter originates from an extreme form of homophily \cite{mcpherson2001birds}, {\em my friends are similar to me}, which justifies that the original Louvain limits the search to the set of the neighbours of the target node. In the case of signed networks, in contrast, the archetype of community structure is a k-clusterable  network \cite{traag2019partitioning}, i.e. a network that can be decomposed into communities where positive edges are inside the clusters and negative edges are across the clusters. K-clusterable graphs are known to exhibit the weak version of structural balance (when $k=2$, they exhibit the strong version). For this reason, when searching for nodes that might be in the same cluster than the target node, it is natural to search among the enemies of its enemies, in the negative layer, as those may be similar to the target node.
For these reasons, {Signed Louvain} is defined with $d_{+}=1$ and $d_{-}=2$, thus restricting nodes to move either in the communities of their positive neighbours ($d_{+}=1$) or to join communities of the nodes that are 2 steps away in the negative layer ($d_{-}=2$).  As part of our experimentation, we also consider an extended version of {Signed Louvain} defined for $d_{+}=d_{-}=2$\footnote{Note that despite the fact that $d_{+}$ and $d_{-}$ take the same value, here $2$, the exploration in the positive and in the negative layer are expected to be very different, due to the differences in their structure. The network of positive connections is expected to present a high clustering coefficient, making the exploration in two steps very similar to that in one step. In contrast, the network of negative connections is expected to have a  small clustering  coefficient \cite{szell2010multirelational}, making the exploration in two steps move away from the direct neighbourhood of the target node.},
thus allowing nodes to move in any community within 2 hops on either layer. The value $d_{+}=2$ allows to consider a bigger set of transitions during the optimisation, and is expected to favour nodes that are in the same triangles as the target node, as those would be simultaneously one and two steps away from the target node. By definition, for $d_{+}=d_{-}=1$, the classical Louvain algorithm is recovered. 

The Signed Louvain algorithm proceeds as follows. The first step is to build the $d_{+}$ and $d_{-}$ networks. In these networks, a node $i$ is neighbour of all the nodes within $d_{+}$ hops or $d_{-}$ hops in the positive and negative layers of the original network respectively. The second step is made of two sub-phases (\textit{move} and \textit{aggregate}) repeated sequentially, similar to the original Louvain algorithm. In the \textit{move} sub-phase, we consider moving each node into the community of any of its neigbours in the $d_{+}$ and $d_{-}$ networks, as long as it achieves a positive gain in modularity. The node is then placed in the community achieving the maximal positive modularity gain. If no positive gain is possible, the node stays in its original community. The process
is applied repeatedly across all nodes until no further local improvement can be achieved. In the \textit{aggregate} phase, a new two-layer network whose nodes are the communities found during the \textit{move} phase is built. The weights of the edges between the new nodes are the sum of the weights of the edges between nodes in the corresponding two communities in each layer of the previous network. The resulting network is therefore also a network with a positive layer and a negative layer. Edges between nodes of the same community
lead to self-loops for this community in the new network. The two sub-phases are repeated until no positive modularity gain is achieved.

The slowest part of the algorithm is the construction of the $d_{+}$- and $d_{-}$-hop graphs in the first step as it requires to scope all $d_{+}$- and $d_{-}$-hop paths starting at every single node of the graph. That being said, by construction, nodes cannot be attached to communities of nodes they cannot reach via a network path. This property is an advantage of {SignedLouvain} over RelaxedLouvain. 

\subsection{Experiments} \label{subsec:signedlouvain_experiments}

In this Section, we demonstrate that {SignedLouvain} exhibits a high quality of optimisation while being computationally much faster than RelaxedLouvain. 

On the synthetic experiment presented in Figure~\ref{fig:LvRLvSL}, we observe that SignedLouvain has a similar performance as RelaxedLouvain. It is able to overcome the over-reliance of the Louvain algorithm on positive edges by exploiting negative edges as well as positive edges. Both SignedLouvain and RelaxedLouvain exhibit a drop of performance for very sparse networks.

As a next step, we compare the modularity and duration achieved by the methods on real-world networks. We note $L$ for Louvain, $RL$ for RelaxedLouvain, $SL_d$ the default version of {SignedLouvain} ($d_{+}=1$, $d_{-}=2$) and $SL_e$ its extended version ($d_{+}=d_{-}=2$). We experiment with DEBAGREEMENT~\cite{pougue2021debagreement}, a set of five networks each extracted from a Reddit forum: BlackLivesMatter, Brexit, Climate, Democrats, Republican. We also consider other topic-specific networks from  BirdwatchSG and TwitterSG, two Twitter networks~\cite{pougue2023learning}. We retain the COVID-19 network in BirdwatchSG. From TwitterSG, we retrieve the networks related to Joel Embiid\footnote{\url{https://en.wikipedia.org/wiki/Joel_Embiid}} (2023 NBA Most Valuable Player), Simone Biles\footnote{\url{https://en.wikipedia.org/wiki/Simone_Biles}} (23-time Gymnastics World Champion and 4-time Olympics Winner between 2013 and 2023), Zinedine Zidane\footnote{\url{https://en.wikipedia.org/wiki/Zinedine_Zidane}} (1998 Football \textit{Ballon d'Or} and 3-time UEFA Champions League Winner as a coach), and Novak Djokovic\footnote{\url{https://en.wikipedia.org/wiki/Novak_Djokovic}} (24-time Tennis Grand Slam Winner).

\begin{figure}
\includegraphics[width = \textwidth]{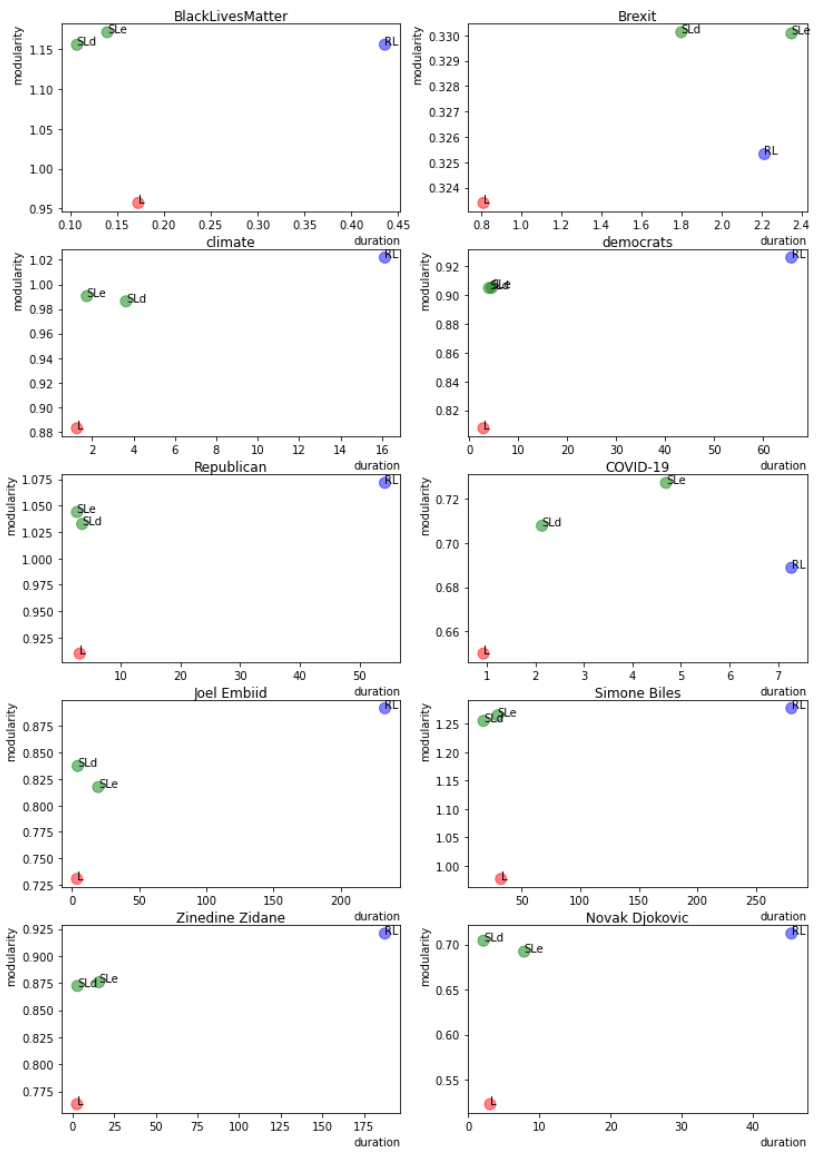}
\caption{Modularity vs Duration. $L$ stands for Louvain, $SL_d$ for default {SignedLouvain}, $SL_e$ for extended {SignedLouvain}, and $RL$ for RelaxedLouvain.}
\label{fig:QvD}
\end{figure}

\begin{table}
\centering
{\renewcommand{\arraystretch}{.9}
\begin{tabular}{| c |c |c |c |c| c|} \hline
 &\textbf{nodes} & \textbf{edges (+\%)}  & \textbf{density} & \makecell{\textbf{average} \\ \textbf{distance}} &  \textbf{diameter}\\  \hline
 \textbf{BLM} & 459 & 461 (58\%) & $10^{-3}$ & 12 & 34  \\ 
 \textbf{Brexit} & 717 & 8486 (42\%) & $10^{-2}$ & 2.4 & 5 \\ 
\textbf{Climate}  & 2694  & 3476 (43\%) & $10^{-3}$ & 5 & 18 \\ 
 \textbf{Democrats}  & 4782 & 6749 (54\%) & $10^{-3}$ & 5 & 15 \\ 
\textbf{Republican}  & 5051 & 5985 (47\%) & $10^{-3}$ & 7 & 20 \\ \hline
\textbf{Joel Embiid}  & 10640 & 15180 (93\%) & $10^{-4}$ & 5 & 17 \\ 
\textbf{Simon Biles}  & 12494 & 13362 (61\%) & $10^{-4}$ & 6  & 23 \\ 
\textbf{Zinedine Zidane} & 9399 & 12554 (96\%)& $10^{-4}$ & 5 & 19 \\ 
\textbf{Novak Djokovic} & 4080 & 8044 (85\%)  & $10^{-3}$ & 5 & 17 \\ \hline
\textbf{Covid-19}  & 1682 & 6648 (66\%) & $10^{-2}$ & 3 & 7 \\ \hline
\end{tabular}}
\caption{List of empirical networks used for the numerical experiments, with some of their key structural properties.} 
\label{tab:dataset_stats}
\end{table}
Figure~\ref{fig:QvD} depicts the result of the experiments. Each algorithm is ran 10 times on each network. The distance and modularity are averaged across the 10 iterations. We observe that {SignedLouvain} is consistently faster than RelaxedLouvain while also matching or improving its modularity. Speed-wise, {SignedLouvain} is competitive with Louvain in most cases. Only on Brexit, $SL_d$ and $SL_e$ are closely on-par with RelaxedLouvain in terms of speed. Across the 10 graphs, {SignedLouvain} is 22 times faster than RelaxedLouvain in average, with a modularity drop of 2\% only. We observe that the relative speed of RelaxedLouvain is sensitive to the size and diameter of the graph. Only for Brexit it is able to compete with the other methods speed-wise. Because the average pair distance is only 2.4, communities are reachable within the 2-hop neighbourhood of any node, which make it equivalent to {SignedLouvain} speed-wise. For all the others graphs, RelaxedLouvain is significantly faster, while providing a  similar performance on most graphs in terms of accuracy. Taken together, these results show that the values $d_{+}=1$ and $d_{-}=2$ provide a good compromise between speed and accuracy, which sugegsts to use them as a default choice when optimising the modularity of signed networks. 
\section{Discussion}

In this article, we have introduced a modification of the Louvain method specifically designed for the clustering of signed networks. When performing a greedy optimisation for community detection, it is essential to explore efficiently the space of node partitions. In the case of Louvain, this search requires one to identify, for each node, a set of nodes that are likely to be in the same community. This set should be as small as possible, to avoid unnecessary operations, but not too small, to avoid missing optimal moves. The construction of the set should ideally be guided by the underlying structure of the graph, and the type of large-scale structure that one aims to uncover. 
As we have discussed, the archetype of community structure for signed networks is associated to the notion of structural balance, suggesting that similar nodes should be two steps away, in the case of negative edges, and one step away, in the case of positive edges. Building on this idea, already at the core of embedding networks for signed networks \cite{kunegis2010spectral,babul2024strong}, we have proposed and successfully tested a variation of the Louvain method, called SignedLouvain. These results open several  directions for future research. First, here we have  considered a modification of the original version of the Louvain method. Yet, several improvements have been developed over the years \cite{blondel2023fast}, most notably the so-called Leiden method \cite{traag2019louvain}, where our results could be incorporated.  Second, an efficient exploration of the space of partitions is critical for modularity optimisation but also, in general, for any optimisation-based method for community detection. As the Louvain method has been adapted to other quality functions, including the Map Equation \cite{rosvall2009map}, it would be interesting to test our search strategy for alternative quality functions designed for signed networks. 
Finally, our results can be interpreted within the language of multiplex networks \cite{kivela2014multilayer}, where nodes would search eligible neighbours differently in different layers - here with two layers, one associated to positive edges and one to negative edges. Generalising our results to the case of more than two layers could improve existing algorithms for the partitioning of multiplex networks \cite{mucha2010community}, which could find direct applications for the mining of networks with complex weights \cite{bottcher2024complex,tian2024structural}.

\section{Acknowledgements}
R.L. acknowledges support from the EPSRC grants EP/V013068/1, EP/V03474X/1 and EP/Y028872/1.

\section{Data and code availability}
Numerical experiments are run on empirical data available through their original publications \cite{pougue2021debagreement,pougue2023learning}. 

The code for this study is available at:
https://github.com/lejohnyjohn/signed-louvain 

\printbibliography

\end{document}